\documentclass[12pt]{article}
\usepackage{graphicx}
\usepackage[bookmarksnumbered,colorlinks,plainpages]{hyperref}

\begin{document}

\title{Action principle and Jaynes' guess method}

\author{Q.A. Wang\\
{\it Institut Sup\'erieur des Mat\'eriaux et M\'ecaniques Avanc\'es}, \\
{\it 44, Avenue F.A. Bartholdi, 72000 Le Mans, France}}

\date{}

\maketitle

\begin{abstract}
A path information is defined in connection with the probability distribution of
paths of nonequilibrium hamiltonian systems moving in phase space from an initial
cell to different final cells. On the basis of the assumption that these paths are
physically characterized by their action, we show that the maximum path information
leads to an exponential probability distribution of action which implies that the
most probable paths are just the paths of stationary action. We also show that the
averaged (over initial conditions) path information between an initial cell and all
the possible final cells can be related to the entropy change defined with natural
invariant measures for dynamical systems. Hence the principle of maximum path
information suggests maximum entropy and entropy change which, in other words, is
just an application of the action principle of classical mechanics to the cases of
stochastic or instable dynamics.
\end{abstract}

\section{Introduction}
The principle of maximum information and entropy are investigated for nonequilibrium
hamiltonian system in connection with action principle of classical mechanics.

The entropy defined by Clausius is an equilibrium quantity\cite{Jaynes} which are
related later by Boltzmann and Gibbs to equilibrium probability distributions with
$S=\ln v$ and $S=-\sum_{i=1}^vp_i\ln p_i$, respectively, here $p_i$ is the
normalized ($\sum_{i=1}^vp_i=1$) probability for the system to be found at state
$i$, $v$ is the total number of states in the phase space occupied by the system
(let Boltzmann constant be unity). The methodology of the statistical mechanics
based on Boltzmann and Gibbs entropies is later complemented by the principle of
maximum entropy\cite{Jaynes,Tribus} which stipulates that the ``best'' equilibrium
probability distributions should maximize Clausius entropy. This method was already
used by Boltzmann and Gibbs\cite{Jaynes,Gibbs}. But thanks to the efforts of
Jaynes\cite{Jaynes2}, it has been nowadays widely recognized to be a powerful
inference method for $guessing$ (not $deriving$) correct probability distributions
of equilibrium as well as nonequilibrium systems. It should be noticed that one of
the underlying philosophical significations of this method is the anthropomorphic
aspect of entropy and information. They are not objective physical quantities. They
are related to the ignorance of the observers. They can be used only for guessing,
but not deriving probability\cite{Jaynes2}.

On the other hand, compared to equilibrium systems, the statistical and
informational methodology for dynamical (nonequilibrium) systems seems less certain.
We have seen recently the extension of the Boltzmann and Gibbs entropies to
nonequilibrium systems (in local equilibrium or not)\cite{Ruelle,Cohen,Lebowitz}.
But as for the inference or guessing methods of nonequilibrium distributions,
although it was assumed by Jaynes that the maximum entropy method always applied,
the principle of minimum entropy production\cite{Prigogine} has been proposed for
the systems sufficiently close to equilibrium state. For certain systems far from
equilibrium, e.g., in the context of the climate of earth, we have still the
principle of maximum entropy production (or minimum entropy
exchange)\cite{Paltridge,Lenton}. Here the entropy is always in the sense of
Clausius, so the entropy production $dS_i\geq 0$ according to the second law. Over
the last years, we have seen the development of a nonextensive statistics which
suggests that the Havrda-Charvat-Daroczy-Tsallis entropy\cite{Esteban} should be
maximized for nonequilibrium systems\cite{Tsallis,Wang02c,Luzzi}, a new extension of
Jaynes method to dynamical systems.

So we have before us many different even opposite methods for dynamical systems. Why
do we have to maximize or minimize entropy or its production for equilibrium as well
as nonequilibrium systems in variational inference method? Are information and
entropy merely anthropomorphic quantities which have nothing to do with physical
laws? Indeed, Jaynes method is not theoretically proved in spite of its success in
``guessing'' probability distributions. It is only supported by plausible arguments
such as, e.g., ``if a system is in a state with lower entropy it would contain more
information than previously specified'', or ``higher entropy states are more
probable'' and ``less predictable'', etc\cite{Jaynes,Jaynes2}. So maximum entropy is
only considered as a principle of guess, but not deduction method of mechanics. The
central quantities of this method, i.e., entropy and information, are only
anthropomorphic\cite{Jaynes2}, not objective quantities involved in fundamental laws
of physics.

In this work, Jaynes approach is revisited under a different angle in connection
with the principle of least (stationary) action. It will be shown that, for
stochastic or instable dynamics of hamiltonian systems, maximum information and
entropy must be used to derive correct probability distributions just as the least
(stationary) action principle must be used for regular dynamics to derive the
correct motions. We finally reach the conclusion that the anthropomorphic aspect of
entropy and information is not needed in order to be able to maximize them for
inference and correct guess. Maximum entropy is in fact a deductive method for
deriving probability distributions. It is an application of the fundamental action
principle of classical mechanics to probabilistic dynamics.

\section{Assumptions and definitions}

Let us begin by some assumptions and definitions concerning dynamical systems.

\begin{enumerate}

\item We recall that a phase space $\Gamma$ of thermodynamic system is defined such
that a physical state of the system is represented by a point in that space. A phase
volume $\Omega$ can be partitioned into $v$ cells of volume $s_i$ with $i=1,2,...v$ in
such a way that $s_i\cap s_j=\emptyset$ ($i\neq j$) and $\cup_{i=1}^vs_i=\Omega$. A
state of the system can be represented by a sufficiently small phase cell in coarse
graining way. The movement of a dynamical system can be represented by its
trajectories (in the sense of classical mechanics) in $\Gamma$ space.

\item The natural invariant measure\cite{Beck} $\mu_i$ is defined as the probability
distribution for a nonequilibrium system to visit different elementary cells $i$ of
its partitioned phase space.

\item The information we address in this work is a measure of the uncertainty of
dynamical process. According to Shannon, this information can be calculated by the
formula
\begin{eqnarray}                                            \label{++2}
H=-\sum_{k=1}^vp_k\ln p_k
\end{eqnarray}
with respect to certain probability distribution $p_k$ of that process and the index
$k$ is summed over all the possible situations.

\item The entropy $S$ of a nonequilibrium system is defined with the natural
invariant measure $\mu_i$ by
\begin{eqnarray}                                            \label{++2+}
S=-\sum_{i=1}^v\mu_i\ln \mu_i.
\end{eqnarray}
If this system reaches equilibrium at the end of a nonequilibrium process and has
$\mu_i$ as its final probability distribution, then $S$ can be considered as the
entropy in the sense of second law (Clausius).
\end{enumerate}

\section{A path information}
In a previous work\cite{Wang04x}, we defined a path information for dynamical
systems moving, over a large period of time, in the $\Gamma$-space between two
points, $a$ and $b$, which are in two elementary cells of a given partition of the
phase space. This path information is given by the Shannon formula with respect to
the probability $p_k(b|a)$ for the system to take the path $k$ from the cell $a$ to
the cell $b$ during $t_{ab}(k)$, as shown in Figure 1(I).

By definition, $p_k(b|a)$ is a transition probability from state $a$ to state $b$
via path $k$. We have $\sum_{k=1}^wp_k(b|a)=1$. This path probability distribution
due to dynamical instability is studied in connection with information theory and
action integral on the basis of the assumption that the different paths are uniquely
differentiated by their actions defined for classical mechanical systems by
\begin{eqnarray}                                            \label{c5}
A_{ab}(k)=\int_{t_{ab}(k)}L_k(t)dt
\end{eqnarray}
where $L_k(t)$ is the Lagrangian of the system at time $t$ along the path $k$. The
average action between state $a$ and state $b$ can be given by
\begin{eqnarray}                                            \label{c5c}
A_{ab}=\sum_{k=1}^wp_k(b|a)A_{ab}(k)
\end{eqnarray}

For all stochastic process like Brownian motion\cite{Kubo}, it can be
proved\cite{Wang04x} that the most probable paths are just the paths of least
action. The probability of other paths decreases exponentially with increasing
action.

In this work, we introduce another approach in order to study a more general
situation where the systems under consideration moves from a initial cell $a$ to
different destinations $b$ with a given travelling time, as shown in Figure 1(II).
This ``spacial uncertainty'' and the ``time uncertainty'' studied in \cite{Wang04x}
are two aspects of the same dynamical instability. That is, for regular motion,
these two uncertainties both disappear.

Now let us consider an ensemble of large $N$ identical systems leaving the initial
cell $a$ in the initial phase volume $A$ for some destinations in the phase volume
$B$ formed by the final phase points occupied by the systems. The travelling time is
$t_{ab}=t_b-t_a$ fixed for every trajectory and destination. After $t_{ab}$, all the
phase points occupied by the systems are found in the volume $B$ partitioned into
cells labelled by $b$. We observe $N_{k_b}$ systems travelling along a path $k_b$
leading to certain cell $b$. A path probability can be defined by
$p_{k_b|a}=N_{k_b}/N$ which is normalized by
\begin{eqnarray}                                            \label{c1xx}
\sum_{b}^{over B}\sum_{k_b=1}^{w_b}p_{k_b|a}=1.
\end{eqnarray}
where $w_b$ is the number of possible paths from $a$ to a given cell $b$ of the
volume $B$. We always suppose each path is characterized by its action $A_{k_b|a}$.
Then an average action can be defined by
\begin{eqnarray}                                            \label{c1xxx}
A_a=\sum_{b}^{over B}\sum_{k_b=1}^{w_b}p_{k_b|a}A_{k_b|a}
\end{eqnarray}
The uncertainty concerning the choice of paths and final cells is measured by the
following Shannon information
\begin{eqnarray}                                            \label{c1x}
H_{B|a}=-\sum_{b}^{over B}\sum_{k_b}p_{k_b|a}\ln p_{k_b|a}
\end{eqnarray}
which can be maximized under the constraints associated with Eq.(\ref{c1xx}) and
Eq.(\ref{c1xxx}) as follows

\begin{eqnarray}                                            \label{xc1x}
\delta (H_{B|a}+\alpha\sum_{b}^{over
B}\sum_{k_b=1}^{w_b}p_{k_b|a}+\eta\sum_{b}^{over
B}\sum_{k_b=1}^{w_b}p_{k_b|a}A_{k_b|a})=0.
\end{eqnarray}
This leads to
\begin{eqnarray}                                            \label{c6x}
p_{k_b|a}=\frac{1}{Z}\exp[-\eta A_{k_b|a}],
\end{eqnarray}
where $Z=\sum_{b}^{over B}\sum_{k_b}\exp[-\eta A_{k_b|a}]$. Putting Eq.(\ref{c6x})
back into Eq.(\ref{c1x}), we get
\begin{eqnarray}                                            \label{c7}
H_{B|a}=\ln Z+\eta A_{a}=\ln Z-\eta \frac{\partial}{\partial\eta}\ln Z.
\end{eqnarray}

It is proved that\cite{Wang04x} the distribution Eq.(\ref{c6x}) is stable with
respect to the fluctuation of action. It is also proved that if $\eta$ is positive,
Eq.(\ref{c6x}) is a least action distribution, i.e., the most probable paths are
just the paths of least action. On the contrary, if $\eta$ is negative, then
Eq.(\ref{c6x}) is a most action distribution which means that the most probable
paths maximize action. In any case, whatever the sign of the parameter $\eta$, the
most probable paths which maximize the path information always correspond to
extremum or stationary action ($\delta A_{k_b|a}=0$). {\it In other words, for
instable dynamical process, the method of maximum information must be used in order
to derive correct probability distribution just as the principle of stationary
action must be used to derive the correct trajectories for regular dynamics}.

In our previous work\cite{Wang04x}, Brownian motion was presented as an example of
the case of $\eta>0$. We would like to mention in passing that this model of
Brownian motion allowed us to give a simple derivation of Fick law for particle
diffusion and of Fourrier law for heat conduction in solids\cite{Wang04y} thanks to
the distribution function Eq.(\ref{c6x}).

\section{Maximum entropy change}
The question here is how to derive the invariant measures $\mu_i$ with which the
entropy $S$ for dynamical systems is defined through Eq.(\ref{++2+}).

In what follows, we suppose that entropy changes when a system moves from the initial
cell $a$ to the final cell $b$. The initial entropy of the systems at time $t=0$ is
given by Shannon formula:
\begin{eqnarray}                                            \label{x1}
S_a=-\sum_{a}\mu_a\ln\mu_a
\end{eqnarray}
where $\mu_a$ is the invariant measure of the systems in the initial phase volume
$A$ at $t=0$ and the cell index $a$ is summed over all the occupied cells in $A$.
After a period of time $t_{ab}$, the system travelling along a path $k_b$ is found
in a cell $b$ with probability measure $\mu_{k_b}(a)$. Supposing the transition
probability is independent of the distribution $\mu_a$ of initial conditions, we
have
\begin{eqnarray}                                            \label{x2}
\mu_{k_b}(a)=\mu_ap_{k_b|a}.
\end{eqnarray}
The entropy of the system after $t_{ab}$ should be defined by
\begin{eqnarray}                                            \label{x3}
S_b &=& -\sum_{a}^{over A}\sum_{b}^{over B}\sum_{k_b}\mu_{k_b}(a)\ln \mu_{k_b}(a)\\
\nonumber &=& -\sum_{a}^{over A}\mu_a\ln \mu_a
-\sum_{a}^{over A}\mu_a\sum_{b}^{over B}\sum_{k_b}p_{k_b|a}\ln p_{k_b|a} \\
\nonumber &=& S_a+\sum_a\mu_aH_{B|a}=S_a+H_{B|A}.
\end{eqnarray}
where $H_{B|A}$ is the average of the path information $H_{b|a}$ over all the
initial phase volume $A$. So if $H_{B|a}$ is maximized, $H_{B|A}$ should also be
maximized. The sums in the above equation must be over all the cells in the initial
phase volume $A$, in the final phase volume $B$ (formed by all cells $b$) and over
all the possible paths between $A$ and $B$.

Eq.(\ref{x3}) can be written as
\begin{eqnarray}                                            \label{x3x}
\Delta S=S_b-S_a=H_{B|A}=\langle\ln Z\rangle+\eta\langle A\rangle,
\end{eqnarray}
where $\langle\ln Z\rangle$ and $\langle A\rangle$ are the averages of $\ln Z$ and
$A_a$ on the volume $A$. Eq.(\ref{x3x}) is the main result of this work. It is
obvious that the maximization of $H_{B|A}$ means the maximization of the entropy
change $\Delta S$ of the dynamical process.

\section{What about maximum entropy?}
Maximum entropy change for dynamical systems does not exclude the possibility of
maximum entropy. In fact, if the entropy change is maximized at any moment of an
evolution with respect to different probability distributions, given the initial
entropy $S_a$, the entropy $S_b$ of the final states must also be at maximum all the
time with respect to the same distributions. So in order to derive correct
probabilities or invariant measures, both entropy change and entropy can be
maximized, depending on which of the two quantities is available as function of the
probability distributions to be derived.

If the initial phase volume $A$ and the final volume $B$ represent equilibrium
states, i.e., the dynamical process connects two Clausius entropies $S_a$ and $S_b$,
then Boltzmann and Gibbs formula can be maximized to derive equilibrium probability
distributions $p_i$ of final states, as suggested by Jaynes principle. This can be
considered as a derivation of the maximum entropy method from the maximum path
information which is nothing but an application of the action principle of classical
mechanics.

\section{Concluding remarks}
A path information is defined in connection with different possible paths of
dynamical system moving in its phase space from the initial cell towards different
final cells. On the basis of the assumption that the paths are physically
differentiated by their actions, we show that the maximum path information leads to
a path probability distribution implying that the most probable paths are just the
paths of stationary action. It is also shown that the average path information
defined in this work is just the variation of the entropy during the dynamical
process. Hence the principles of least action and of maximum path information
suggest the maximum entropy change for dynamical systems. For given initial
distribution, maximum entropy change means in fact the maximization of final entropy
with respect to the same probability distributions.

We would like to mention that the intrinsic link between entropy and action tells us
that the information theory and statistical mechanics do not need that entropy be an
anthropomorphic quantity. Information and entropy are nothing but objective measures
of dynamical uncertainty when the motion of physical systems become probabilistic.
So maximum information and entropy is not merely an inference principle. It is a law
of physics. This is the main conclusion of this paper.

\newpage

\begin{figure}[ht] \label{f1}
\caption{(I) The possible phase space paths ($k=1,2...w$) for a system to go from
the cell $a$ to the cell $b$ in time $t_{ab}(k)$, each having a probability
$P_{ab}(k)$. If $w$ is the number of all the possible paths, we have
$\sum_{k=1}^wp_{ab}(k)=1$. (II) The possible phase space paths for a system to go
from a given cell $a$ to the different cells $b$ of a partition of the final phase
volume $B$ during the time $t_{ab}$, each having a probability $p_{k_b|a}$. More
than one paths from $a$ to a given cell $b$ are allowed so we may have $k_b=1,2 ...
w_b$ where $w_b$ is the total number of paths from $a$ to a cell $b$.}
\end{figure}

\newpage

\includegraphics[width=12cm,height=16cm]{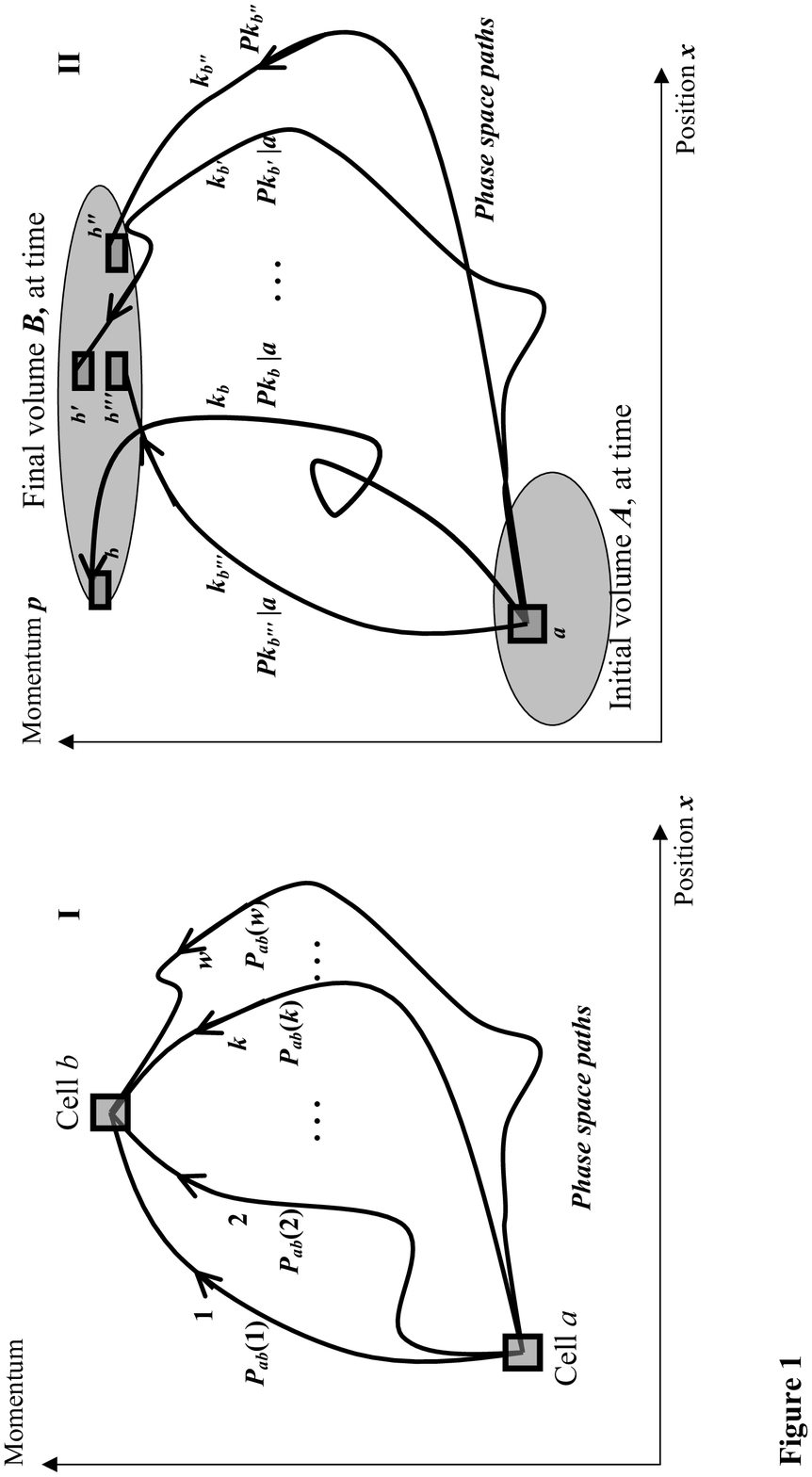}


\begin{thebibliography}{99}

\bibitem {Jaynes}
E.T. Jaynes, {\em The evolution of Carnot's principle}, The opening talk at the EMBO
Workshop on Maximum Entropy Methods in x-ray crystallographic and biological
macromolecule structure determination, Orsay, France, April 24-28, 1984; Reprinted
in Ericksen $\&$ Smith, Vol 1, pp. 267-282; and {\em Phys. Rev.,} {\bf
106},620(1957)

\bibitem {Tribus}
M. Tribus, {\em D\'ecisions Rationelles dans l'incertain,} (Paris, Masson et Cie,
1972)P14-26; or {\em Rational, descriptions, decisions and designs,} (Pergamon Press
Inc., 1969)

\bibitem {Gibbs}
J. Willard Gibbs, {\em Principes \'el\'ementaires de m\'ecanique statistique,}
(Paris, Hermann, 1998)

\bibitem {Ruelle}
D. Ruelle, Is there a unified theory of nonequilibrium statistical mechanics?
Proceedings of the International Conference on Theoretical Physics Th-2002 (Paris,
July 22-27, 2002), Birkh\"auser Verlag, Berlin, 2004, p.489;

D. Ruelle, Extending the definition of entropy to nonequilibrium steady state, {\em
Proc. Nat. Acad. Sci.,} {\bf 100},30054(2003)


\bibitem {Cohen}
G. Gallavotti, E.G.D. Cohen, {\em Note on nonequilibrium stationary states and
entropy,} cond-mat/0312306

\bibitem {Lebowitz}
P.L. Garrido, S. Goldstein and J.L. Lebowitz, {\em The Boltzmann entropy for dense
fluid not in local equilibrium,} cond-mat/0310575

S. Goldstein and J.L. Lebowitz, {\em On the (Boltzmann) Entropy of Nonequilibrium
Systems,} cond-mat/0304251

\bibitem {Prigogine}
I. Prigogine, {\em Bull. Roy. Belg. Cl. Sci.,} {\bf 31},600(1945)

\bibitem {Paltridge}
G. Paltridge, {\em Quart. J. Roy. Meteor. Soc.,} {\bf 101},475(1975)

\bibitem {Lenton}
Tim Lenton, {\em Maximum entropy production in Edinburgh, a report of the second
'Daisyworld and beyond' workshop,} (2002), document obtained from
\url{http://www.cogs.susx.ac.uk/daisyworld/ws2002_overview.html}

\bibitem {Esteban}
M.D. Esteban, {\em Kybernetika,\/} {\bf 31}(1995)337

J. Harvda and F. Charvat, {\em Kybernetika,\/} {\bf 3}(1967)30

Z. Daroczy, {\em Information and control,\/} {\bf 16}(1970)36

\bibitem {Tsallis}
C. Tsallis, {\em J. Stat. Phys.,} {\bf 52},479(1988); C. Tsallis, F. Baldovin, R.
Cerbino and P. Pierobon, {\em Introduction to Nonextensive Statistical Mechanics and
Thermodynamics}, cond-mat/0309093

\bibitem {Wang02c}
Q.A. Wang, {\em Euro. Phys. J. B,} {\bf26}(2002)357

\bibitem {Luzzi}
R. Luzzi, A.R. Vasconcellos, J.G.Ramos, Entropy: Mystery and Controversy. Plethora
of Informational-Entropies and Unconventional Statistics, cond-mat/0307325v4

\bibitem {Jaynes2}
E.T. Jaynes, Gibbs vs Boltzmann entropies, {\em American Journal of Physics}, {\bf
33},391(1965);

E.T. Jaynes, Where do we go from here? {\em Maximum entropy and Bayesian methods in
inverse problems}, pp.21-58, eddited by C. Ray Smith and W.T. Grandy Jr., D. Reidel
Publishing Company (1985)

\bibitem {Beck}
C. Beck and F. Schl\"ogl, {\em Thermodynamics of chaotic systems}, Cambridge
University Press, 1993

\bibitem {Wang04x}
Q.A. Wang, Maximum path information and the principle of least action for chaotic
system, {\em Chaos, Solitons $\&$ Fractals,} (2004), in press; cond-mat/0405373 and
ccsd-00001549

\bibitem {Kubo}
R. Kubo, M. Toda, N. Hashitsume, {\em Statistical physics II, Nonequilibrium
statistical mechanics,\/} Springer, Berlin, 1995

\bibitem {Wang04y}
Q.A. Wang, Maximum entropy change and least action principle for nonequilibrium
systems, Invited talk at the Twelfth United Nations/European Space Agency Workshop
on Basic Space Science, 24-28 May 2004, Beijing, P.R. China; cond-mat/0312329

\end{thebibliography}
\end{document}